\documentclass[fleqn,usenatbib]{mnras}

\usepackage{newtxtext,newtxmath}

\usepackage[T1]{fontenc}
\usepackage{ae,aecompl}

\usepackage{graphicx}	
\usepackage{amsmath}	
\usepackage{multicol}        
\usepackage{bm}		
\usepackage{pdflscape}	
\usepackage{url}

\title[Age-metallicity trends in the MW disc]{Chasing the impact of the Gaia-Sausage-Enceladus merger on the formation of the Milky Way thick disc
}
\author[I. Ciuc\u{a} et al.]{
Ioana Ciuc\u{a}$^{1,2, 3}$,\thanks{E-mail: ioana.ciuca@anu.edu.au}
Daisuke Kawata$^{4}$,
Yuan-Sen Ting$^{1,2}$,
Robert J. J. Grand$^{5, 6}$,
\newauthor{
Andrea Miglio$^{7, 8}$,
Michael Hayden$^{3, 9}$,
Junichi Baba$^{10, 11}$,
Francesca Fragkoudi$^{12}$,}
\newauthor{
Stephanie Monty$^{13}$,
Sven Buder$^{1, 3}$,
Ken Freeman$^{1}$
} 
\\
$^{1}$Research School of Astronomy \& Astrophysics, Australian National University, Canberra, ACT 2611, Australia \\
$^{2}$School of Computing, Australian National University, Canberra, ACT 2601, Australia \\
$^{3}$ARC Centre of Excellence for All Sky Astrophysics in 3 Dimensions (ASTRO 3D), Australia \\
$^{4}$Mullard Space Science Laboratory, University College London,	Holmbury St. Mary, Dorking, Surrey, RH5 6NT, UK \\
$^{5}$Instituto de Astrofísica de Canarias, Calle Via Lactea s/n, E-38205 La Laguna, Tenerife, Spain \\
$^{6}$Departamento de Astrof\'isica, Universidad de La Laguna, Av. del Astrof\'isico Francisco S\'anchez s/n, E-38206, La Laguna,Tenerife, Spain \\
$^{7}$Dipartimento di Fisica e Astronomia Augusto Righi, Università degli Studi di Bologna, Via Gobetti 93/2, I-40129 Bologna, Italy \\
$^{8}$INAF–Osservatorio di Astrofisica e Scienza dello Spazio, Via P. Gobetti 93/3, 40129 Bologna, Italy \\
$^{9}$Sydney Institute for Astronomy, School of Physics, A28, The University of Sydney, NSW, 2006, Australia \\
$^{10}$National Astronomical Observatory of Japan, Mitaka-shi, Tokyo 181-8588, Japan \\
$^{11}$Department of Astronomical Science, SOKENDAI (The Graduate University of Advanced Studies), Mitaka, Tokyo 181-8588, Japan \\
$^{12}$Institute for Computational Cosmology, Department of Physics, Durham University, Durham DH1 3LE, UK \\
$^{13}$Institute of Astronomy, University of Cambridge, Madingley Road, Cambridge, CB3 0HA, UK}

\date{Accepted XXX. Received YYY; in original form ZZZ}

\pubyear{2020}

\begin{document}
\label{firstpage}
\pagerange{\pageref{firstpage}--\pageref{lastpage}}
\maketitle

\begin{abstract}
We employ our Bayesian Machine Learning framework BINGO (Bayesian INference for Galactic archaeOlogy) to obtain high-quality stellar age estimates for 68,360 red giant and red clump stars present in the $17^{\rm th}$ data release of the Sloan Digital Sky Survey, the APOGEE-2 high-resolution spectroscopic survey. By examining the denoised age-metallicity relationship of the Galactic disc stars, we identify a drop in metallicity with an increase in [Mg/Fe] at an early epoch, followed by a chemical enrichment episode with increasing [Fe/H] and decreasing [Mg/Fe]. This result is congruent with the chemical evolution induced by an early-epoch gas-rich merger identified in the Milky Way-like zoom-in cosmological simulation Auriga. In the initial phase of the merger of Auriga 18 there is a drop in metallicity due to the merger diluting the metal content and an increase in the [Mg/Fe] of the primary galaxy. Our findings suggest that the last massive merger of our Galaxy, the Gaia-Sausage-Enceladus, was likely a significant gas-rich merger and induced a starburst, contributing to the chemical enrichment and building of the metal-rich part of the thick disc at an early epoch. 
\end{abstract}




\section{Introduction}
\label{sec:intro}
Our Galaxy has an entangled history that challenges our understanding of its formation. Early in its evolution, it experienced a Galactic-altering event known as the Gaia-Sausage-Enceladus (GSE) merger \citep{Belokurov_2018, Helmi_2018}, which recast its chemical and dynamical make-up. The extent of its impact is now well-established thanks to the new kinematic information extracted from ESA's flagship astrometric survey \textit{Gaia} \citep{gaia2016a} and the high-quality spectroscopic data coming from surveys such as APOGEE \citep{Maj2017}, GALAH \citep{Buder_2021}, Gaia-ESO \citep{Gilmore_2012} or LAMOST \citep{Zhao_2012}.

The consensus is that the GSE was the last massive merger ($M_{\star} \simeq 10^{9}$ $\rm{M_{\odot}}$) of the Milky Way and happened early ($8-11$ Gyr ago) \citep[e.g.,][]{Vin2019, Bel_2020}. It greatly impacted the Galactic halo, delivering approximately two-thirds of its stellar component in the form of stars on highly-eccentric orbits \citep[e.g.][]{Mackereth+Bovy20}. These stars inhabit a sausage-like distribution in the radial-azimuthal velocity distribution \citep{Brook_2003, Belokurov_2018} and the blue sequence of the Hertzsprung–Russell diagram of the halo stars. GSE members also appear to be more metal-poor and less $\alpha$-enhanced than the redder halo counterpart \citep{Haywood_2018a, Helmi_2018}. The origin of these red-sequence stars referred to as the Splash by \cite{Bel_2020}, but initially found by other earlier studies \citep{DiMatteo+Haywood+Lehnert+19, Gallart+Bernard+Brook+19}, is thought to be the result of proto-galactic disc stars being dynamically ejected into the halo during the GSE merger.

\begin{figure}
\centering
 \includegraphics[width=0.8\hsize]{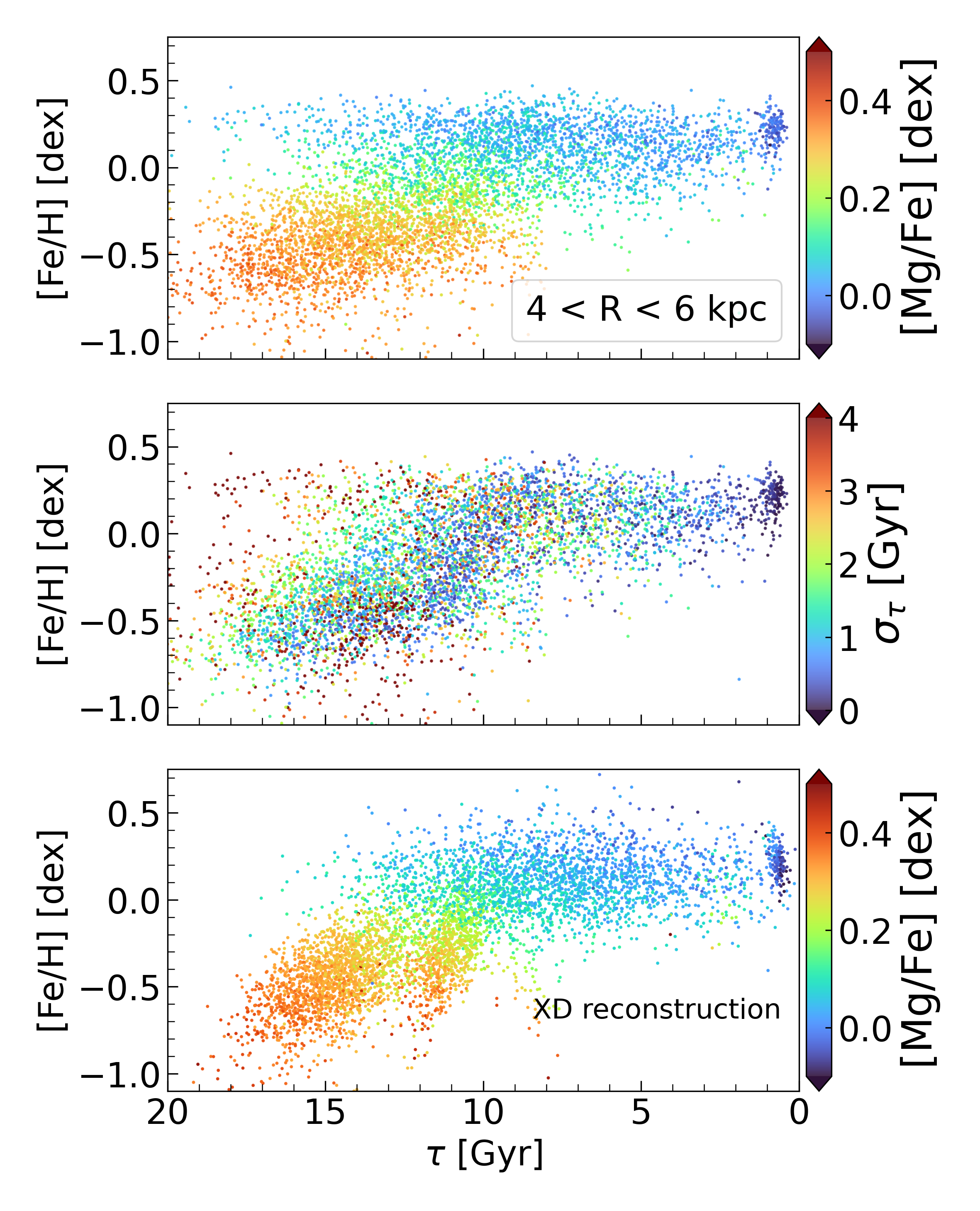}
\caption{The age-metallicity relation for the stars within $4 < R < 6$~kpc. The top panel displays the original distribution of the age and [Fe/H] for stars coloured by $\rm [Mg/Fe]$. The middle panel shows the same population of stars as the top panel but is coloured by the age uncertainty, $\sigma_{\tau}$. The bottom panel shows the age-metallicity and [Mg/Fe] relation reconstructed with an extreme deconvolution algorithm (see Section~\ref{sec:method}), which recovers the noise-free probability distribution spanned by the age, metallicity and [Mg/Fe] after deconvolving the noisy distribution with a measurement error model assumed to be Gaussian.   
}
\label{fig:proof_xd}
\end{figure}
The certainty of GSE's existence, supported by its indubitable imprint on the stellar halo, demands a thorough investigation of its impact on the chemo-dynamical structure of the proto-galactic disc. By analysing simulated galaxies with a thick disc component, \cite{Brook_2004, Brook_2006} posed a formation scenario for thick discs, in which they form during an early time characterised by intense hierarchical merging. The gas-rich mergers trigger an episode of intense star formation, during which most canonical thick disc stars form. This scenario is supported by the recent work of \cite{Grand_2020}, who analysed the Auriga suite of cosmological zoom-in Milky Way-like simulations. Their results support the heavy blueprint of the GSE-like merger on the formation of the thick disc, which has a two-fold effect. Firstly, the GSE merger heats a part of the existent proto-disc stars, which dynamically ejects them into the halo and creates the Splash. Secondly, it brings fresh gas into the central galactic regions, which triggers a starburst that forms the stars of a younger thick disc. The thin disc then begins forming post-merger from the gradual accretion of metal-poor gas in an inside-out, upside-down fashion \citep[e.g.,][]{Bird2013, Grand2018}. Studying the stellar population around the solar radius of the APOGEE DR14 data, \citet{Ciuca_2021} presented a qualitatively consistent observational trend with this scenario. However, further testing the validity of such scenarios using observed data requires a representative sample of disc stars with reliable stellar ages. 

In this \textit{Letter}, we examine the age-metallicity relationship (AMR) for a sample of giant stars in the APOGEE-2 Data Release 17 (DR17) survey \citep{Abd_2021}, with ages and their uncertainties determined using the robust Bayesian INference for Galactic archaeOlogy approach \citep[BINGO,][]{Ciuca_2021}. In Section~\ref{sec:method}, we describe BINGO and the scalable extreme deconvolution technique used to denoise the age, metallicity and the [Mg/Fe] abundance distribution. Section~\ref{sec:results} reveals the denoised AMR across the full extent of the Milky Way disc and discusses the results, considering the insights from the cosmological simulations of the Milky Way-like galaxies.

\begin{figure}
\centering
 \includegraphics[width=\hsize]{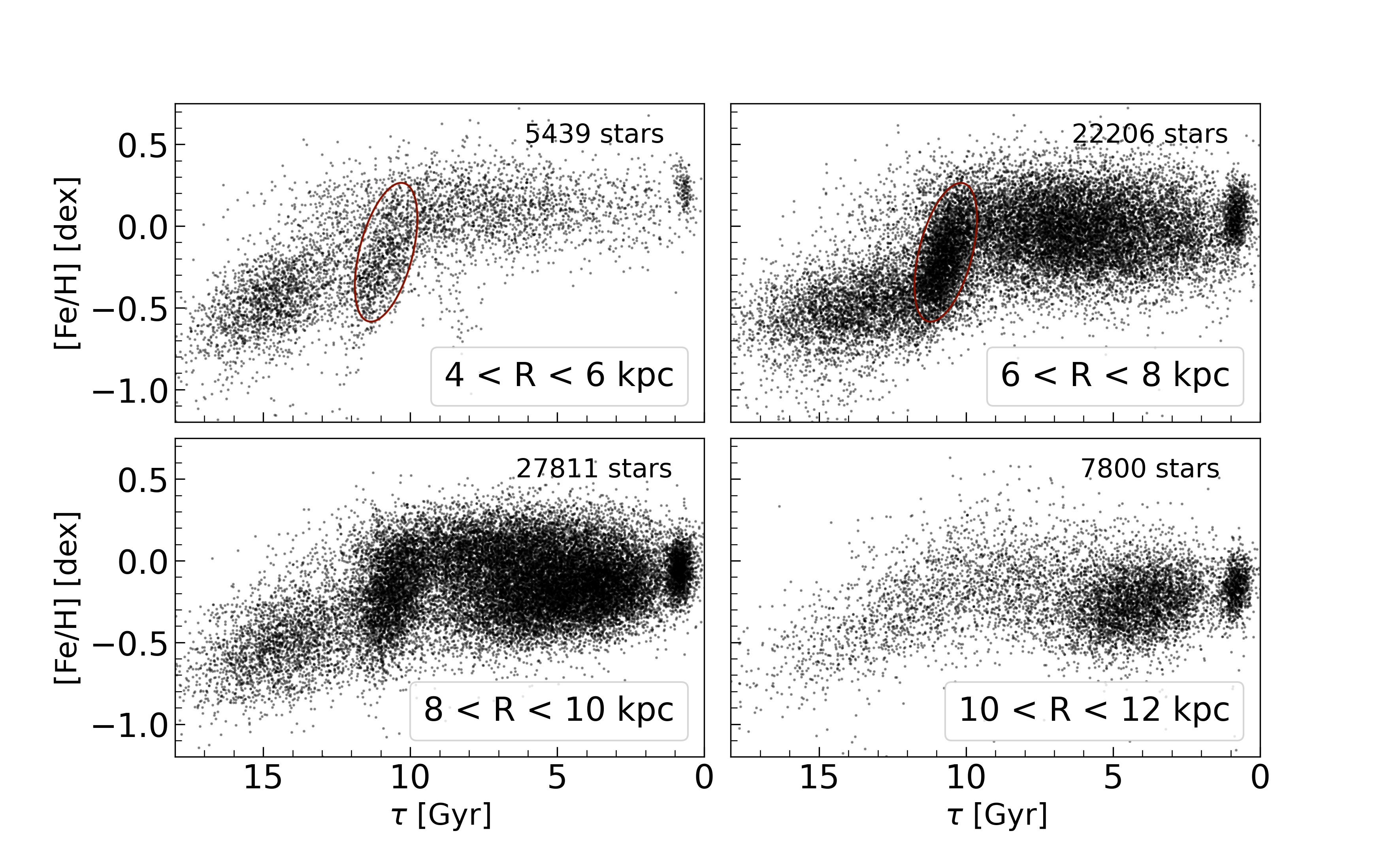}
\caption{The denoised distribution of metallicity as a function of age, $\tau$, across the radial extent of the Galactic disc. Ellipses in the top panels highlight the GGS (see text).
}
\label{fig:amr_nocol}
\end{figure}

\begin{figure*}
\centering
 \includegraphics[width=0.7\textwidth]{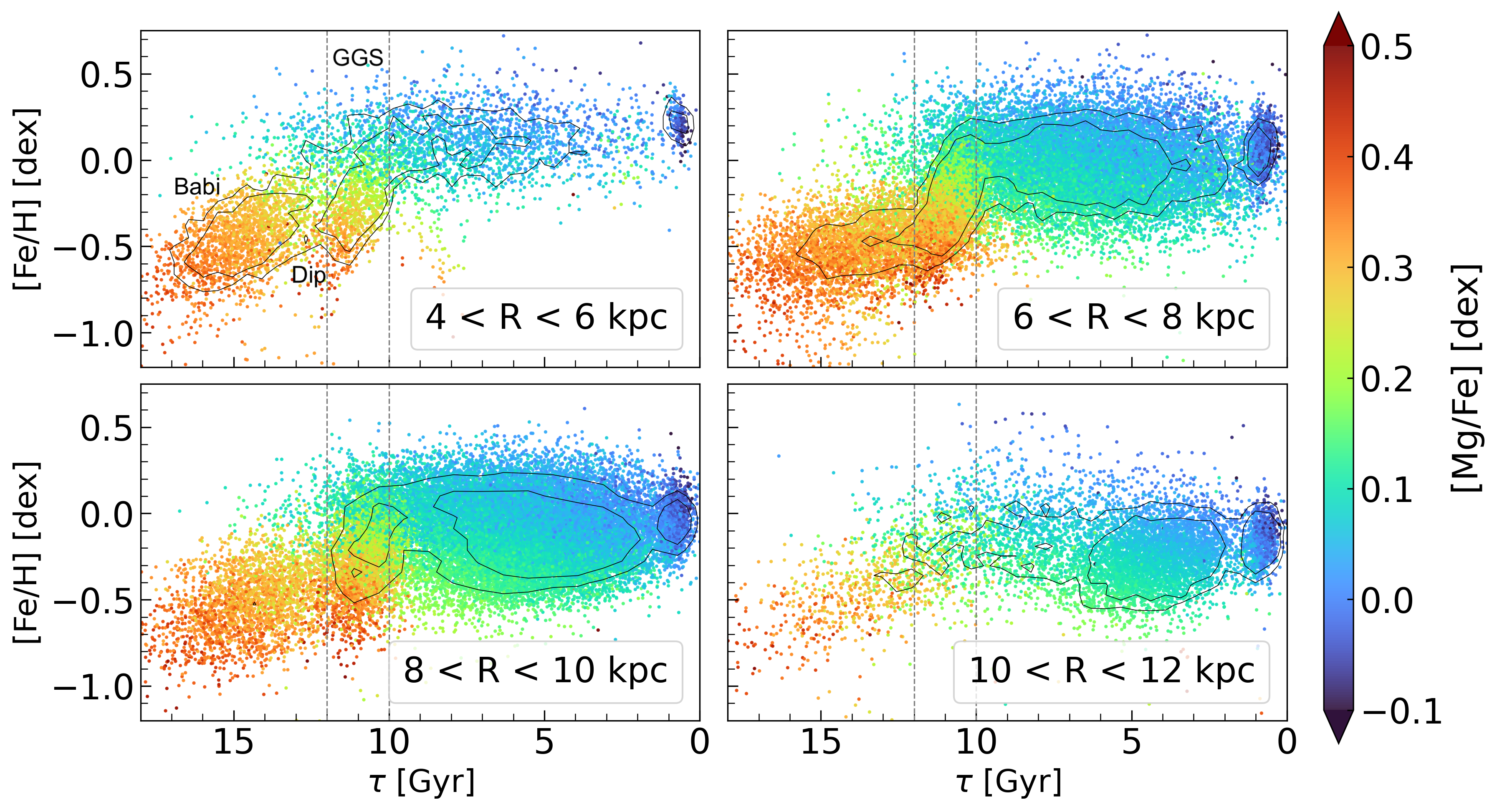}
\caption{The AMR across the entire radial extent for our XD resampled sample of stars, from the most inner region ($4 < R < 6$~kpc) in the upper left panel to the outer disc ($10 < R < 12$~kpc) in the lower right panel. The colour of the dots indicates $\rm [Mg/Fe]$. In the inner region, we designate the key features: the older population Babi, the Dip and the GGS. Overlaid are the contours enclosing 50  and 75 \% of the population.
}
\label{fig:amr_all}
\end{figure*}

\section{Data and Method}
\label{sec:method}
Following the same method as in \cite{Ciuca_2021}, we construct a Bayesian Neural Network to map the stellar parameters, T$_{\rm eff}$, $\log g$, [Fe/H] and the [Mg/Fe], [C/Fe] and [N/Fe] abundances together with their associated uncertainties to the asteroseismic ages derived by \citet{Miglio_2020} for the APOKASC-2 stars \citep{Pins2018}. 
To use only the highest quality data and the tracer stars with a reliable asteroseismic age, from the data of \cite{Miglio_2020}, we select red clump (RC) stars with a mass higher than 1.8~$\rm M_{\odot}$ and red giant branch (RGB) stars, with high signal-to-noise (SNR) greater than 100 in the associated APOGEE-2 spectrum.

We employ the BINGO model only on a set of stellar data that traces the same population as the training data, i.e. to a specific population of RC stars with a mass higher than $1.8 \ M_{\odot}$ or RGB stars, as done in \cite{Ciuca_2021}. To this end, we train a neural network model built using {\tt Keras} and {\tt TensorFlow} \citep{Abadi_2016} on the original APOKASC-2 data to classify RC stars with a mass higher than $1.8 \ M_{\odot}$ and RGB stars. Our strategy is similar to that used in \cite{Ting_2018} to identify RC stars. 

We then apply the trained classifier to the sample of APOGEE-2 stars with SNR $>100$, $T_{\rm eff}$ between 4,000 and 5,500~K, and $\log g$  between 1 and 3.5, to be within the training set limits of BINGO. We only select stars with a probability higher than 0.95 of being classified as RGB or high-mass RC stars. Finally, we remove the duplicates of the spectra for the same star by only retrieving the highest SNR ones.

We are left with 89,591 stars, for which we obtain the age estimates by applying BINGO on their stellar parameters and the [Mg/Fe], [C/Fe] and [N/Fe] abundances. To further ensure the quality of our sample, we remove the data with an age uncertainty of $\sigma_{\log \tau} > 0.2$. We further discard the stars younger than 8~Gyr that have [Mg/Fe] $>0.2$~dex, which we deem to be merged binary stars, i.e. artificially younger stars \citep[e.g.][]{Aguirre_2018, Ciuca_2021}. Our final sample contains 68,360 stars. We employ the Galactic radius, $R$, which we compute from the recommended distance in the \texttt{astroNN} \citep{Leung+Bovy19b} catalogue of APOGEE~DR17, assuming a solar radius of $R_0=8$~kpc.

Using the data described above, we analyse the AMR of the stars in the different Galactic radial, $R$, bins. Fig.~\ref{fig:proof_xd} shows the AMR for our stars within $4 < R < 6$~kpc. Note that the ages of some of the old stars are much older than the age of the Universe. This is because \cite{Miglio_2020} used a highly uninformative prior on the maximum age of 40~Gyr that reflects in our asteroseismic age measurement \citep{Miglio_2020}. This is deliberate since using a strong prior could compress the old end of the AMR and potentially obscure the GSE signature. We consider that our age estimate is reliable in terms of relative age. The absolute age scale is only indicative but follows the asteroseismic age scale in \citet{Miglio_2020}. Also, note that the sizable number of stars younger than $\sim1.5$~Gyr is due to our cut of the low-mass RC stars.
 
The top panel of Fig.~\ref{fig:proof_xd} displays the original distribution of the age and [Fe/H] for stars coloured by [Mg/Fe]. We notice the presence of a high-metallicity ([Fe/H] $>0$), old ($\tau>13$~Gyr) population with low [Mg/Fe] $<0.1$, which is intriguing given that we expect low [Mg/Fe] to be associated with a younger population. However, the middle panel shows that these stars have a large uncertainty (up to 4~Gyr in linear age) in their age estimates, and the existence of such stars is not statistically significant. Here, we define the linear age uncertainty of $\sigma_{\tau}$ as $0.5 (10^{\log \tau + \sigma_{\log \tau}}-10^{\log \tau - \sigma_{\log \tau}})$, because BINGO estimates the age of the stars in $\log$.

\begin{figure}
\centering
 \includegraphics[width=\hsize]{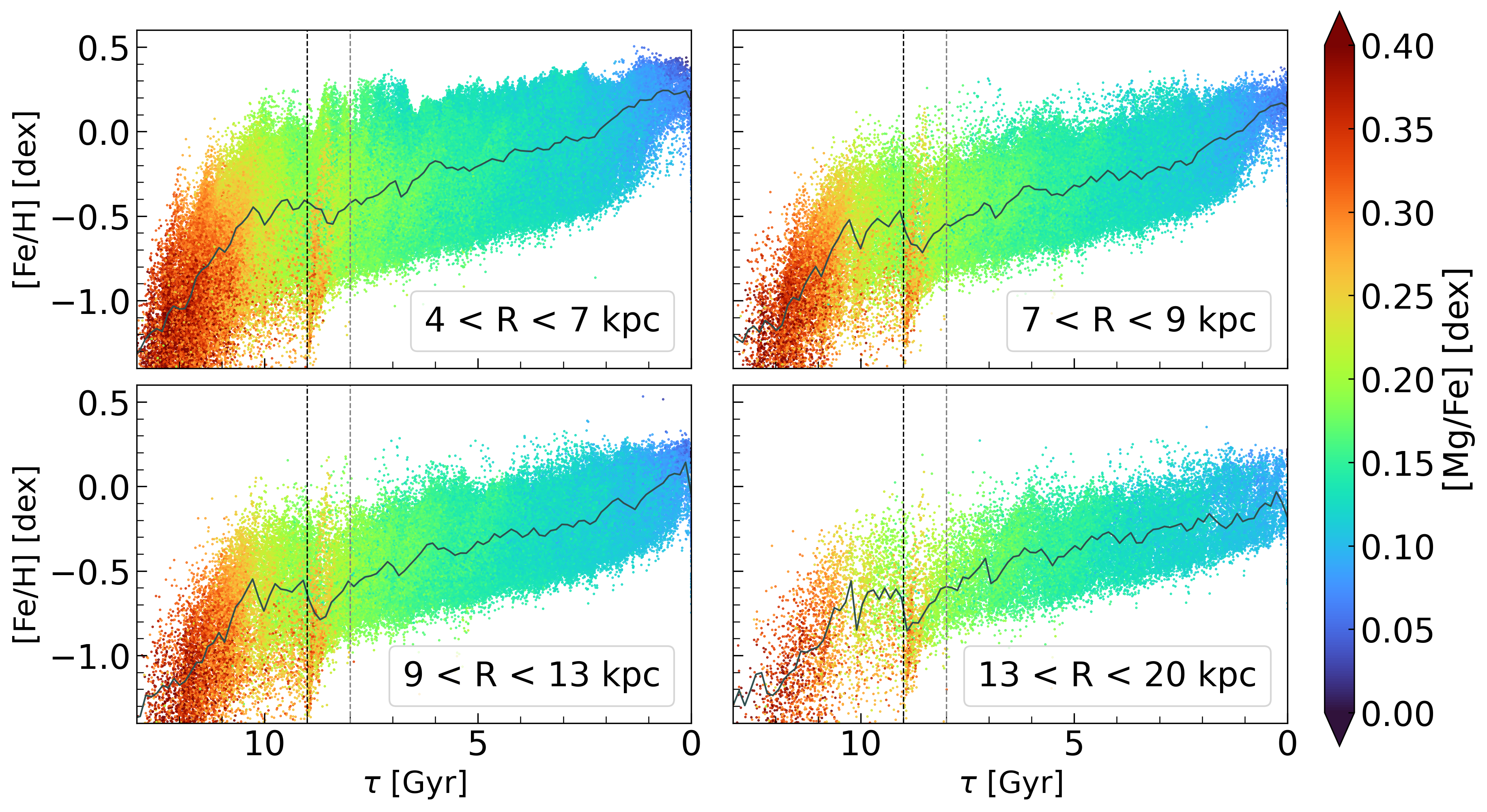}
\caption{The AMR coloured by [Mg/Fe] across the radial extent of the Milky Way-like Au18 zoom-in cosmological simulation. The vertical dotted lines highlight the period of the Au18 gas-rich merger.}
\label{fig:au18}
\end{figure}

To eliminate the spurious features due to the lower confidence data and highlight the statistically  significant trends only, we employ scalable extreme deconvolution \citep[XD,][]{Ritchie+Murray19}. We model the three-dimensional distribution of age, [Fe/H] and [Mg/Fe] as a Gaussian Mixture Model (GMM), considering all the uncertainties. We perform density deconvolution with a 15-component GMM to recover the denoised distribution in the inner radial bin of $4 < R < 6$~kpc, and we use ten components for the other radial bins. Our choice of the number of components for the GMM ensures a low training error per radial bin. The bottom panel of Fig.~\ref{fig:proof_xd} shows the reconstructed distribution of age, [Fe/H] and [Mg/Fe] from the GMM. XD clears the spurious features from the high uncertainty stars, e.g. high-metallicity and very old stars, which reassures us that XD performs well with this task. Unless otherwise stated, we present the results of the AMR reconstructed with XD, i.e. XD-denoised AMR.

\section{Results and Discussion}
\label{sec:results}

Fig.~\ref{fig:amr_nocol} shows the XD-denoised AMR from our stellar sample within four different radial bins, $4 < R < 6$~kpc, $6 < R < 8$~kpc, $8 < R < 10$~kpc and $10 < R < 12$~kpc. For the inner radial bin, we removed all stars with $R < 4$~kpc to minimise contamination from the bulge. Note that we do not employ any data cut using the height from the disc midplane or kinematics. Instead, our results in this \textit{Letter} focus on the stars with [Fe/H] $>-1$, i.e. studying the thick and thin disc regime in the Milky Way.

The upper panels, in particular the upper-left panel of $4 < R < 6$~kpc, reveal an old population ($\tau > 13$~Gyr) with a low, but similar metallicity as the thick disc population ([Fe/H] $\simeq-0.5$), which we denote Babi\footnote{The word `Babi' comes from the Romanian language and denotes a much-cherished grandmother.}. This panel displays that the metallicity around the age of $\tau=13$~Gyr, [Fe/H] $\simeq-0.3$, goes down to [Fe/H] $\simeq-0.5$ for the younger age of around $\tau\simeq12$~Gyr, which we refer to as the Dip. This drop in metallicity is followed by an increase of $\rm [Fe/H]$ from $\tau\simeq12$~Gyr and $\rm [Fe/H] \simeq -0.5$~dex to $\tau\simeq10$ Gyr and $\rm [Fe/H] \simeq 0.2$~dex, i.e., the diagonal blob-like phase, which we denote the Great Galactic Starburst or GGS phase. The Dip and GGS features are also present in the lower-left panel of Fig.~\ref{fig:amr_nocol}, although the features are clearer in the inner disc. We consider that the Babi and Dip likely correspond to the Thick Disc I and II populations, which were chemically identified in \citet{Anders+Chiappini+Santiago+18}.

\begin{figure}
\centering
 \includegraphics[width=\hsize]{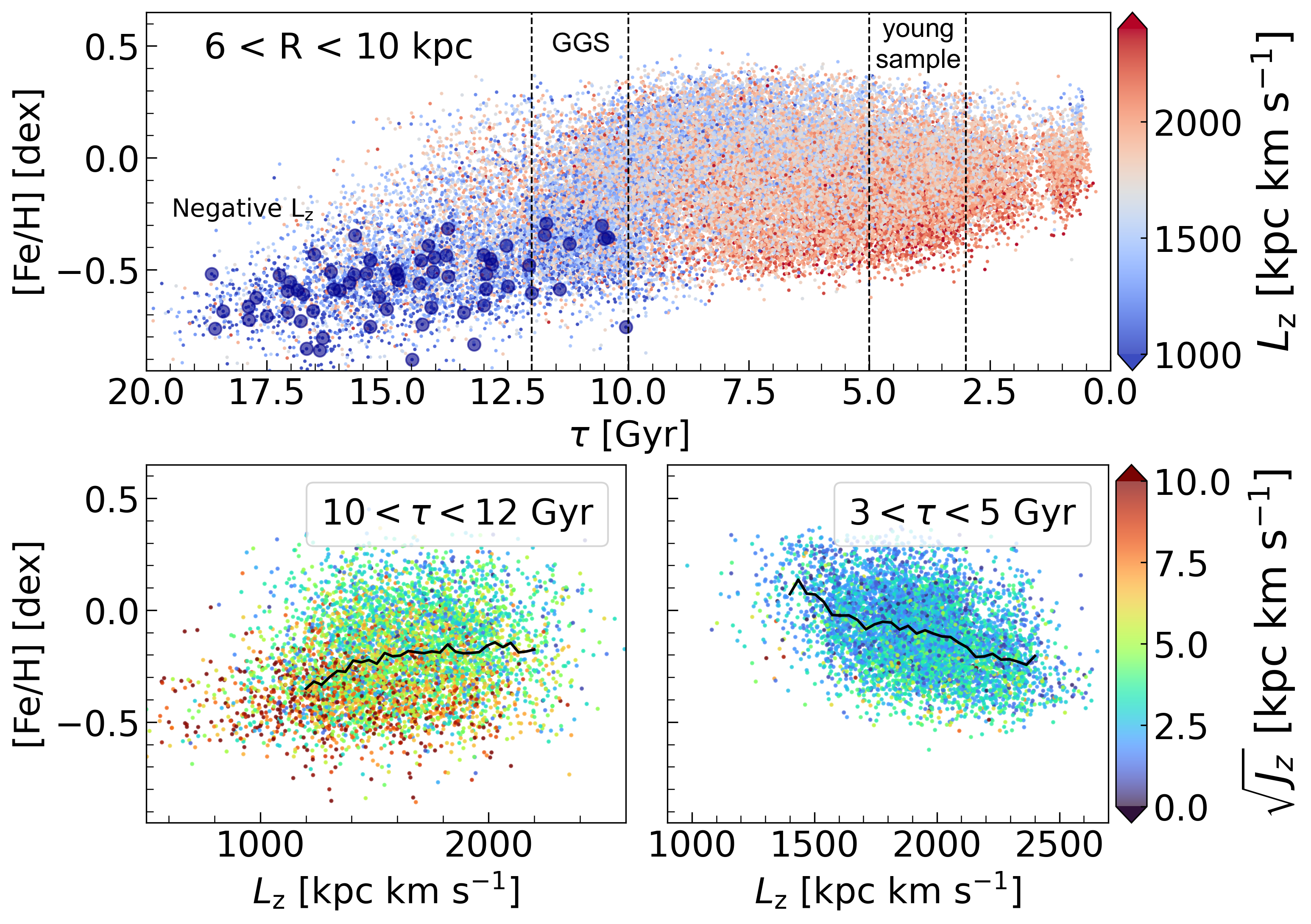}
\caption{The noisy AMR coloured by angular momentum, $L_\textrm{z}$ (Top panel), where we highlight the stars with negative angular momentum as dark blue dots. The lower left panel shows $L_\textrm{z}-$[Fe/H] relation for the stars with ages between 10 and 12~Gyr, i.e. the GGS phase. The solid line represents the median value of [Fe/H] as a function of $L_\textrm{z}$. The colour shows the square root of the vertical action, $J_\textrm{z}$. The lower right panel shows the same as the bottom left panel, but for the stars with ages between 3 and 5~Gyr, i.e. young, thin disc phase. The vertical dotted lines in the upper panel highlight the age ranges of the stars in the lower panels.}
\label{fig:lz_gradient}
\end{figure}

Fig.~\ref{fig:amr_all} shows the same AMR as in Fig.~\ref{fig:amr_nocol}, but colour coded with $\rm [Mg/Fe]$. We remark that the Dip feature of the drop in [Fe/H], which we identified at $\tau\simeq12$~Gyr, is accompanied by an increase in [Mg/Fe]. Contrastingly, the diagonal GGS shows that [Mg/Fe] decreases with increasing [Fe/H] from $\tau\simeq12$~Gyr to $\tau\simeq10$~Gyr. 

Strikingly, we find similar features in one of the Auriga simulations, Auriga~18 (Au18). Auriga is a series of zoom-in cosmological simulations of the Milky Way-like disc galaxies \citep{Grand2017Au}. Fig.~\ref{fig:au18} shows the AMR colour coded with [Mg/Fe] for the star particles in Au18 within four different radial bins. We used different radial bins because the Au18 disc is larger than the Milky Way disc, and we discard the star particles within $R<4$~kpc since our APOGEE-2 data have very few stars in the very inner disc ($R\leq3$~kpc). 

Especially in the panels of $R>7$~kpc, we can see the drop of [Fe/H] around the age of $\tau\simeq9$~Gyr. Au18 shows a significant gas-rich merger around that time \citep{Grand+Kawata+Belokurov+20}. The vertical dotted lines highlight the period of a starburst induced by a gas-rich merger, as seen in Fig.~2 of \citet{Grand+Kawata+Belokurov+20}. The drop in [Fe/H] coincides with the beginning of the gas-rich merger. Interestingly, the lower [Fe/H] stars that formed at the beginning of the merger show higher [Mg/Fe] than the older stars, in agreement with the observational data of the Dip shown in Fig.~\ref{fig:amr_all}. Similarly to the GGS feature in the observational data, the Dip feature in Au18 precedes the diagonal feature of increasing [Fe/H] and decreasing [Mg/Fe] for younger stars. 

As mentioned above, these features coincide with the gas-rich merger of Au18. As shown in \citet{Bustamante+Sparre+Springel+Grand18} based on the Auriga simulation data, the gas-rich merger can bring the low metallicity gas and dilute the metallicity, and the fresh enrichment from the star formation induced by the merger can drive the higher [Mg/Fe] \citep[see also][]{Brook+Richard+Kawata+07}. Then, as indicated in \citet{Grand+Kawata+Belokurov+20}, the gas-rich merger induces a starburst, which leads to the chemical enrichment that explains the increase of [Fe/H] apace with the decrease of [Mg/Fe] soon after the peak of the starburst \citep{Brook+Richard+Kawata+07}, as seen in the diagonal GGS feature. \citet{Grand+Kawata+Belokurov+20}, who focuses on this particular merger in Au18, also suggest that the gas-rich merger also induced the generation of the relatively metal-rich part of the thick disc. 

Hence, the observed Dip and GGS features in Fig.~\ref{fig:amr_all} can be explained as the impact of the most significant merger. This is close to the epoch of the Milky Way's last significant merger, the GSE merger \citep[e.g.][]{Montalban+Mackereth+Miglio+21}. Therefore, these trends in Fig.~\ref{fig:amr_all} could indicate that the GSE merger was a significant gas-rich merger and that the gas brought by the GSE drove the metallicity lower. This was followed by significant star formation during the merger, which enriched the Galaxy. It is interesting to see that although a similar GGS feature is seen in the $4 < R < 6$~kpc bin, the AMR is more populated in metal-poor older stars. This may indicate that the gas-rich merger-induced starburst of the GSE started in the inner disc, which is consistent with the suggested radially-plunging orbit of the GSE merger \citep[e.g.,][]{Vasiliev_2022}. Also, the similarity between the distribution of [Fe/H] and [Mg/Fe] as a function of age in the GGS among the different radial bins in Fig.~\ref{fig:amr_all} suggests that the chemical evolution in this merger phase is relatively well-mixed, which suggests that the star formation in this period happened in a relatively compact region. This is consistent with the small radial size of the high-[Mg/Fe] thick disc stars.

We further examine how the kinematics of the stars changes across the disc's temporal evolution. For this purpose, we construct the AMR using only high-confidence age data, i.e. $\sigma_{\log \tau}<0.05$ corresponding to $\sigma_{\tau} \ \lesssim 1.5$~Gyr for the $\tau=10$~Gyr stars.  Fig.~\ref{fig:amr_all} shows that, for the population of stars younger than 8~Gyr, there is a lack of lower [Fe/H] stars in the $R<6$~kpc bin, while for $10<R<12 $~kpc, the trend is reversed, with the high [Fe/H] stars missing. Owing to radial migration, the stars in $6<R<10$~kpc include both high and low metallicity stars formed in the inner and outer disc, respectively. Therefore, we present the stellar data solely within $6<R<10$~kpc. We employ the angular momentum, $L_\mathrm{z}$, and the vertical action, $J_\mathrm{z}$, for these stars, taken from the \texttt{astroNN} catalogue of APOGEE~DR17.

The upper panel of  Fig.~\ref{fig:lz_gradient} shows the AMR of the high-age confidence stars coloured with $L_\mathrm{z}$. As expected from the well-established negative radial metallicity gradient, the stars younger than $\simeq8$~Gyr show the higher $L_\mathrm{z}$ for lower [Fe/H] at a fixed age population. Interestingly, the higher [Fe/H] stars show higher $L_\mathrm{z}$ during the GGS phase.
Also, for a fixed [Fe/H] population, the younger stars have higher $L_\mathrm{z}$, which indicates the younger stars formed in the outer disk and/or the younger stars have colder kinematics. 

In the lower panel of Fig.~\ref{fig:lz_gradient}, we show the $L_\mathrm{z}-$[Fe/H] relation for the thin disc population whose ages are between 3 and 5~Gyr (lower right panel) and the GGS phase, i.e. stars with $10<\tau<12$~Gyr (lower right panel), and we use the colour to denote  $\sqrt{J_\mathrm{z}}$. The younger thin disc displays the well-known trend of higher [Fe/H] in the inner disc, i.e. lower $L_\mathrm{z}$. Also, higher $L_\mathrm{z}$ stars with lower metallicity have higher $J_\mathrm{z}$, i.e. vertically flaring \citep[e.g.][]{Rahimi+Carrell+Kawata14,Kawata_2017}. Conversely, the younger thick-disc population shows a positive gradient of the $L_\mathrm{z}-$[Fe/H] relation, as traced by the median trend indicated by the solid line, which is opposite to the thin disc \citep{Spagna+Lattanzi+Fiorentin10, Lee+Beers+Ivezic+11} and can be interpreted as the inside-out formation of the thick disc \citep{Schoenrich+McMillan17, Kawata_2018}. The panel also shows that the lower $L_\mathrm{z}$ stars with lower [Fe/H] have higher $J_\mathrm{z}$, which support the upside-down formation of the thick disc \citep[e.g.][]{Brook_2004,Bird2013,Kawata_2018,Ciuca_2021}. Hence, this trend indicates that during the gas-rich merger, the thick disc becomes kinematically colder as [Fe/H] increases and [Mg/Fe] decreases. 

In addition, there appears to be a lack of metal-poor stars ([Fe/H] $\leq0$) with ages $\tau\sim 9-10$~Gyr compared to stars younger than $\sim9$ Gyr in the upper panel of Fig.~\ref{fig:lz_gradient}. This tentative feature suggests that the chaotic phase of thick disc formation ended around 9 billion years ago. After this time, the thin disc began to grow and establish the negative metallicity gradient, as pointed out as the Bridge phase in \cite{Ciuca_2021}.

In the top panel of Fig.~\ref{fig:lz_gradient}, we highlight the stars with negative angular momentum as dark blue dots, which occupy the predominantly older age regime of the AMR, i.e., Babi, in agreement with \cite{Gallart+Bernard+Brook+19} and \cite{Xiang_2022}. These counter-rotating stars with [Fe/H] $\simeq-0.5$~dex are likely to be the Splash component due to the GSE merger \citep{DiMatteo+Haywood+Lehnert+19, Belokurov_2018}.  The kinematics of Babi and GGS in Fig.~\ref{fig:lz_gradient} indicates their similarity to Pops. D and C in \cite{Sahl_2022}. GGS and Babi likely trace the Pops C and the metal-rich end of Pop. D, respectively. While \cite{Sahl_2022} considers Pops D and C to overlap, Babi and GGS formed before and during the GSE merger.

The precise relative age inferred for the APOGEE-2 stars enables us to identify further evidence of the significant gas-rich merger of GSE. Our results suggest that the GSE significantly impacted the formation of the Galactic disc, and it catalysed the transition from the thick disc to the thin disc formation as suggested in \cite{Grand+Kawata+Belokurov+20} and \cite{Ciuca_2021}. 

\section*{Data Availability}
The data underlying this article will be shared on reasonable request to the corresponding author.

\section*{Acknowledgements}

IC is grateful for the Joint Jubilee Fellowship at the Australian National University. DK acknowledges the support of the UK's Science and Technology Facilities Council (STFC Grant ST/S000216/1 and ST/W001136/1 ).  AM acknowledges support from the ERC Consolidator Grant funding scheme (project ASTEROCHRONOMETRY, grant agreement number 772293). RG has received funding from the European Research Council (ERC) under the European Union's Horizon 2020 research and innovation programme (CartographY GA. 804752). RG acknowledges financial support from the Spanish Ministry of Science and Innovation (MICINN) through the Spanish State Research Agency under the Severo Ochoa Program 2020-2023 (CEX2019-000920-S). JB acknowledges the support by JSPS KAKENHI grant Nos. 21K03633 and 21H00054.
%



\bibliographystyle{mnras}
\bibliography{./icms}

\begin{thebibliography}{}
\makeatletter
\relax
\def\mn@urlcharsother{\let\do\@makeother \do\$\do\&\do\#\do\^\do\_\do\%\do\~}
\def\mn@doi{\begingroup\mn@urlcharsother \@ifnextchar [ {\mn@doi@}
  {\mn@doi@[]}}
\def\mn@doi@[#1]#2{\def\@tempa{#1}\ifx\@tempa\@empty \href
  {http://dx.doi.org/#2} {doi:#2}\else \href {http://dx.doi.org/#2} {#1}\fi
  \endgroup}
\def\mn@eprint#1#2{\mn@eprint@#1:#2::\@nil}
\def\mn@eprint@arXiv#1{\href {http://arxiv.org/abs/#1} {{\tt arXiv:#1}}}
\def\mn@eprint@dblp#1{\href {http://dblp.uni-trier.de/rec/bibtex/#1.xml}
  {dblp:#1}}
\def\mn@eprint@#1:#2:#3:#4\@nil{\def\@tempa {#1}\def\@tempb {#2}\def\@tempc
  {#3}\ifx \@tempc \@empty \let \@tempc \@tempb \let \@tempb \@tempa \fi \ifx
  \@tempb \@empty \def\@tempb {arXiv}\fi \@ifundefined
  {mn@eprint@\@tempb}{\@tempb:\@tempc}{\expandafter \expandafter \csname
  mn@eprint@\@tempb\endcsname \expandafter{\@tempc}}}

\bibitem[\protect\citeauthoryear{{Abadi} et~al.,}{{Abadi}
  et~al.}{2016}]{Abadi_2016}
{Abadi} M.,  et~al., 2016, arXiv e-prints, \href
  {https://ui.adsabs.harvard.edu/abs/2016arXiv160304467A} {p. arXiv:1603.04467}

\bibitem[\protect\citeauthoryear{{Abdurro'uf} et~al.,}{{Abdurro'uf}
  et~al.}{2022}]{Abd_2021}
{Abdurro'uf} et~al., 2022, \mn@doi [\apjs] {10.3847/1538-4365/ac4414}, \href
  {https://ui.adsabs.harvard.edu/abs/2022ApJS..259...35A} {259, 35}

\bibitem[\protect\citeauthoryear{{Anders}, {Chiappini}, {Santiago},
  {Matijevi{\v{c}}}, {Queiroz}, {Steinmetz}  \& {Guiglion}}{{Anders}
  et~al.}{2018}]{Anders+Chiappini+Santiago+18}
{Anders} F.,  {Chiappini} C.,  {Santiago} B.~X.,  {Matijevi{\v{c}}} G.,
  {Queiroz} A.~B.,  {Steinmetz} M.,   {Guiglion} G.,  2018, \mn@doi [\aap]
  {10.1051/0004-6361/201833099}, \href
  {https://ui.adsabs.harvard.edu/abs/2018A&A...619A.125A} {619, A125}

\bibitem[\protect\citeauthoryear{{Belokurov}, {Erkal}, {Evans}, {Koposov}  \&
  {Deason}}{{Belokurov} et~al.}{2018}]{Belokurov_2018}
{Belokurov} V.,  {Erkal} D.,  {Evans} N.~W.,  {Koposov} S.~E.,   {Deason}
  A.~J.,  2018, \mn@doi [\mnras] {10.1093/mnras/sty982}, \href
  {https://ui.adsabs.harvard.edu/abs/2018MNRAS.478..611B} {478, 611}

\bibitem[\protect\citeauthoryear{{Belokurov}, {Sanders}, {Fattahi}, {Smith},
  {Deason}, {Evans}  \& {Grand}}{{Belokurov} et~al.}{2020}]{Bel_2020}
{Belokurov} V.,  {Sanders} J.~L.,  {Fattahi} A.,  {Smith} M.~C.,  {Deason}
  A.~J.,  {Evans} N.~W.,   {Grand} R. J.~J.,  2020, \mn@doi [\mnras]
  {10.1093/mnras/staa876}, \href
  {https://ui.adsabs.harvard.edu/abs/2020MNRAS.494.3880B} {494, 3880}

\bibitem[\protect\citeauthoryear{{Bird}, {Kazantzidis}, {Weinberg}, {Guedes},
  {Callegari}, {Mayer}  \& {Madau}}{{Bird} et~al.}{2013}]{Bird2013}
{Bird} J.~C.,  {Kazantzidis} S.,  {Weinberg} D.~H.,  {Guedes} J.,  {Callegari}
  S.,  {Mayer} L.,   {Madau} P.,  2013, \mn@doi [\apj]
  {10.1088/0004-637X/773/1/43}, \href
  {https://ui.adsabs.harvard.edu/abs/2013ApJ...773...43B} {773, 43}

\bibitem[\protect\citeauthoryear{{Brook}, {Kawata}, {Gibson}  \&
  {Flynn}}{{Brook} et~al.}{2003}]{Brook_2003}
{Brook} C.~B.,  {Kawata} D.,  {Gibson} B.~K.,   {Flynn} C.,  2003, \mn@doi
  [\apjl] {10.1086/374306}, \href
  {https://ui.adsabs.harvard.edu/abs/2003ApJ...585L.125B} {585, L125}

\bibitem[\protect\citeauthoryear{{Brook}, {Kawata}, {Gibson}  \&
  {Freeman}}{{Brook} et~al.}{2004}]{Brook_2004}
{Brook} C.~B.,  {Kawata} D.,  {Gibson} B.~K.,   {Freeman} K.~C.,  2004, \mn@doi
  [\apj] {10.1086/422709}, \href
  {https://ui.adsabs.harvard.edu/abs/2004ApJ...612..894B} {612, 894}

\bibitem[\protect\citeauthoryear{{Brook}, {Kawata}, {Martel}, {Gibson}  \&
  {Bailin}}{{Brook} et~al.}{2006}]{Brook_2006}
{Brook} C.~B.,  {Kawata} D.,  {Martel} H.,  {Gibson} B.~K.,   {Bailin} J.,
  2006, \mn@doi [\apj] {10.1086/499154}, \href
  {https://ui.adsabs.harvard.edu/abs/2006ApJ...639..126B} {639, 126}

\bibitem[\protect\citeauthoryear{{Brook}, {Richard}, {Kawata}, {Martel}  \&
  {Gibson}}{{Brook} et~al.}{2007}]{Brook+Richard+Kawata+07}
{Brook} C.,  {Richard} S.,  {Kawata} D.,  {Martel} H.,   {Gibson} B.~K.,  2007,
  \mn@doi [\apj] {10.1086/511056}, \href
  {https://ui.adsabs.harvard.edu/abs/2007ApJ...658...60B} {658, 60}

\bibitem[\protect\citeauthoryear{{Buder} et~al.,}{{Buder}
  et~al.}{2021}]{Buder_2021}
{Buder} S.,  et~al., 2021, \mn@doi [\mnras] {10.1093/mnras/stab1242}, \href
  {https://ui.adsabs.harvard.edu/abs/2021MNRAS.506..150B} {506, 150}

\bibitem[\protect\citeauthoryear{{Bustamante}, {Sparre}, {Springel}  \&
  {Grand}}{{Bustamante} et~al.}{2018}]{Bustamante+Sparre+Springel+Grand18}
{Bustamante} S.,  {Sparre} M.,  {Springel} V.,   {Grand} R. J.~J.,  2018,
  \mn@doi [\mnras] {10.1093/mnras/sty1692}, \href
  {https://ui.adsabs.harvard.edu/abs/2018MNRAS.479.3381B} {479, 3381}

\bibitem[\protect\citeauthoryear{{Ciuc{\u{a}}}, {Kawata}, {Miglio}, {Davies}
  \& {Grand}}{{Ciuc{\u{a}}} et~al.}{2021}]{Ciuca_2021}
{Ciuc{\u{a}}} I.,  {Kawata} D.,  {Miglio} A.,  {Davies} G.~R.,   {Grand} R.
  J.~J.,  2021, \mn@doi [\mnras] {10.1093/mnras/stab639}, \href
  {https://ui.adsabs.harvard.edu/abs/2021MNRAS.503.2814C} {503, 2814}

\bibitem[\protect\citeauthoryear{{Di Matteo}, {Haywood}, {Lehnert}, {Katz},
  {Khoperskov}, {Snaith}, {G{\'o}mez}  \& {Robichon}}{{Di Matteo}
  et~al.}{2019}]{DiMatteo+Haywood+Lehnert+19}
{Di Matteo} P.,  {Haywood} M.,  {Lehnert} M.~D.,  {Katz} D.,  {Khoperskov} S.,
  {Snaith} O.~N.,  {G{\'o}mez} A.,   {Robichon} N.,  2019, \mn@doi [\aap]
  {10.1051/0004-6361/201834929}, \href
  {https://ui.adsabs.harvard.edu/abs/2019A&A...632A...4D} {632, A4}

\bibitem[\protect\citeauthoryear{{Gaia Collaboration} et~al.,}{{Gaia
  Collaboration} et~al.}{2016}]{gaia2016a}
{Gaia Collaboration} et~al., 2016, \mn@doi [\aap]
  {10.1051/0004-6361/201629512}, \href
  {http://adsabs.harvard.edu/abs/2016A%26A...595A...2G} {595, A2}

\bibitem[\protect\citeauthoryear{{Gallart}, {Bernard}, {Brook}, {Ruiz-Lara},
  {Cassisi}, {Hill}  \& {Monelli}}{{Gallart}
  et~al.}{2019}]{Gallart+Bernard+Brook+19}
{Gallart} C.,  {Bernard} E.~J.,  {Brook} C.~B.,  {Ruiz-Lara} T.,  {Cassisi} S.,
   {Hill} V.,   {Monelli} M.,  2019, \mn@doi [Nature Astronomy]
  {10.1038/s41550-019-0829-5}, \href
  {https://ui.adsabs.harvard.edu/abs/2019NatAs...3..932G} {3, 932}

\bibitem[\protect\citeauthoryear{{Gilmore} et~al.,}{{Gilmore}
  et~al.}{2012}]{Gilmore_2012}
{Gilmore} G.,  et~al., 2012, The Messenger, \href
  {https://ui.adsabs.harvard.edu/abs/2012Msngr.147...25G} {147, 25}

\bibitem[\protect\citeauthoryear{{Grand} et~al.,}{{Grand}
  et~al.}{2017}]{Grand2017Au}
{Grand} R. J.~J.,  et~al., 2017, \mn@doi [\mnras] {10.1093/mnras/stx071}, \href
  {https://ui.adsabs.harvard.edu/abs/2017MNRAS.467..179G} {467, 179}

\bibitem[\protect\citeauthoryear{{Grand} et~al.,}{{Grand}
  et~al.}{2018}]{Grand2018}
{Grand} R. J.~J.,  et~al., 2018, \mn@doi [\mnras] {10.1093/mnras/stx3025},
  \href {https://ui.adsabs.harvard.edu/abs/2018MNRAS.474.3629G} {474, 3629}

\bibitem[\protect\citeauthoryear{{Grand} et~al.,}{{Grand}
  et~al.}{2020a}]{Grand_2020}
{Grand} R. J.~J.,  et~al., 2020a, \mn@doi [\mnras] {10.1093/mnras/staa2057},
  \href {https://ui.adsabs.harvard.edu/abs/2020MNRAS.497.1603G} {497, 1603}

\bibitem[\protect\citeauthoryear{{Grand} et~al.,}{{Grand}
  et~al.}{2020b}]{Grand+Kawata+Belokurov+20}
{Grand} R. J.~J.,  et~al., 2020b, \mn@doi [\mnras] {10.1093/mnras/staa2057},
  \href {https://ui.adsabs.harvard.edu/abs/2020MNRAS.497.1603G} {497, 1603}

\bibitem[\protect\citeauthoryear{{Haywood}, {Di Matteo}, {Lehnert}, {Snaith},
  {Khoperskov}  \& {G{\'o}mez}}{{Haywood} et~al.}{2018}]{Haywood_2018a}
{Haywood} M.,  {Di Matteo} P.,  {Lehnert} M.~D.,  {Snaith} O.,  {Khoperskov}
  S.,   {G{\'o}mez} A.,  2018, \mn@doi [\apj] {10.3847/1538-4357/aad235}, \href
  {https://ui.adsabs.harvard.edu/abs/2018ApJ...863..113H} {863, 113}

\bibitem[\protect\citeauthoryear{Helmi, Babusiaux, Koppelman, Massari,
  Veljanoski  \& Brown}{Helmi et~al.}{2018}]{Helmi_2018}
Helmi A.,  Babusiaux C.,  Koppelman H.~H.,  Massari D.,  Veljanoski J.,   Brown
  A. G.~A.,  2018, \mn@doi [Nature] {10.1038/s41586-018-0625-x}, 563, 85–88

\bibitem[\protect\citeauthoryear{{Kawata}, {Grand}, {Gibson}, {Casagrande},
  {Hunt}  \& {Brook}}{{Kawata} et~al.}{2017}]{Kawata_2017}
{Kawata} D.,  {Grand} R. J.~J.,  {Gibson} B.~K.,  {Casagrande} L.,  {Hunt} J.
  A.~S.,   {Brook} C.~B.,  2017, \mn@doi [\mnras] {10.1093/mnras/stw2363},
  \href {https://ui.adsabs.harvard.edu/abs/2017MNRAS.464..702K} {464, 702}

\bibitem[\protect\citeauthoryear{{Kawata} et~al.,}{{Kawata}
  et~al.}{2018}]{Kawata_2018}
{Kawata} D.,  et~al., 2018, \mn@doi [\mnras] {10.1093/mnras/stx2464}, \href
  {https://ui.adsabs.harvard.edu/abs/2018MNRAS.473..867K} {473, 867}

\bibitem[\protect\citeauthoryear{{Lee} et~al.,}{{Lee}
  et~al.}{2011}]{Lee+Beers+Ivezic+11}
{Lee} Y.~S.,  et~al., 2011, \mn@doi [\apj] {10.1088/0004-637X/738/2/187}, \href
  {https://ui.adsabs.harvard.edu/abs/2011ApJ...738..187L} {738, 187}

\bibitem[\protect\citeauthoryear{{Leung} \& {Bovy}}{{Leung} \&
  {Bovy}}{2019}]{Leung+Bovy19b}
{Leung} H.~W.,  {Bovy} J.,  2019, \mn@doi [\mnras] {10.1093/mnras/stz2245},
  \href {https://ui.adsabs.harvard.edu/abs/2019MNRAS.489.2079L} {489, 2079}

\bibitem[\protect\citeauthoryear{{Mackereth} \& {Bovy}}{{Mackereth} \&
  {Bovy}}{2020}]{Mackereth+Bovy20}
{Mackereth} J.~T.,  {Bovy} J.,  2020, \mn@doi [\mnras] {10.1093/mnras/staa047},
  \href {https://ui.adsabs.harvard.edu/abs/2020MNRAS.492.3631M} {492, 3631}

\bibitem[\protect\citeauthoryear{{Majewski} et~al.,}{{Majewski}
  et~al.}{2017}]{Maj2017}
{Majewski} S.~R.,  et~al., 2017, \mn@doi [\aj] {10.3847/1538-3881/aa784d},
  \href {https://ui.adsabs.harvard.edu/abs/2017AJ....154...94M} {154, 94}

\bibitem[\protect\citeauthoryear{{Miglio} et~al.,}{{Miglio}
  et~al.}{2021}]{Miglio_2020}
{Miglio} A.,  et~al., 2021, \mn@doi [\aap] {10.1051/0004-6361/202038307}, \href
  {https://ui.adsabs.harvard.edu/abs/2021A&A...645A..85M} {645, A85}

\bibitem[\protect\citeauthoryear{{Montalb{\'a}n} et~al.,}{{Montalb{\'a}n}
  et~al.}{2021}]{Montalban+Mackereth+Miglio+21}
{Montalb{\'a}n} J.,  et~al., 2021, \mn@doi [Nature Astronomy]
  {10.1038/s41550-021-01347-7}, \href
  {https://ui.adsabs.harvard.edu/abs/2021NatAs...5..640M} {5, 640}

\bibitem[\protect\citeauthoryear{{Pinsonneault} et~al.,}{{Pinsonneault}
  et~al.}{2018}]{Pins2018}
{Pinsonneault} M.~H.,  et~al., 2018, \mn@doi [\apjs]
  {10.3847/1538-4365/aaebfd}, \href
  {https://ui.adsabs.harvard.edu/abs/2018ApJS..239...32P} {239, 32}

\bibitem[\protect\citeauthoryear{{Rahimi}, {Carrell}  \& {Kawata}}{{Rahimi}
  et~al.}{2014}]{Rahimi+Carrell+Kawata14}
{Rahimi} A.,  {Carrell} K.,   {Kawata} D.,  2014, \mn@doi [Research in
  Astronomy and Astrophysics] {10.1088/1674-4527/14/11/004}, \href
  {https://ui.adsabs.harvard.edu/abs/2014RAA....14.1406R} {14, 1406}

\bibitem[\protect\citeauthoryear{{Ritchie} \& {Murray}}{{Ritchie} \&
  {Murray}}{2019}]{Ritchie+Murray19}
{Ritchie} J.~A.,  {Murray} I.,  2019, arXiv e-prints, \href
  {https://ui.adsabs.harvard.edu/abs/2019arXiv191111663R} {p. arXiv:1911.11663}

\bibitem[\protect\citeauthoryear{{Sahlholdt}, {Feltzing}  \&
  {Feuillet}}{{Sahlholdt} et~al.}{2022}]{Sahl_2022}
{Sahlholdt} C.~L.,  {Feltzing} S.,   {Feuillet} D.~K.,  2022, \mn@doi [\mnras]
  {10.1093/mnras/stab3681}, \href
  {https://ui.adsabs.harvard.edu/abs/2022MNRAS.510.4669S} {510, 4669}

\bibitem[\protect\citeauthoryear{{Sch{\"o}nrich} \& {McMillan}}{{Sch{\"o}nrich}
  \& {McMillan}}{2017}]{Schoenrich+McMillan17}
{Sch{\"o}nrich} R.,  {McMillan} P.~J.,  2017, \mn@doi [\mnras]
  {10.1093/mnras/stx093}, \href
  {https://ui.adsabs.harvard.edu/abs/2017MNRAS.467.1154S} {467, 1154}

\bibitem[\protect\citeauthoryear{{Silva Aguirre} et~al.,}{{Silva Aguirre}
  et~al.}{2018}]{Aguirre_2018}
{Silva Aguirre} V.,  et~al., 2018, \mn@doi [\mnras] {10.1093/mnras/sty150},
  \href {https://ui.adsabs.harvard.edu/abs/2018MNRAS.475.5487S} {475, 5487}

\bibitem[\protect\citeauthoryear{{Spagna}, {Lattanzi}, {Re Fiorentin}  \&
  {Smart}}{{Spagna} et~al.}{2010}]{Spagna+Lattanzi+Fiorentin10}
{Spagna} A.,  {Lattanzi} M.~G.,  {Re Fiorentin} P.,   {Smart} R.~L.,  2010,
  \mn@doi [\aap] {10.1051/0004-6361/200913538}, \href
  {https://ui.adsabs.harvard.edu/abs/2010A&A...510L...4S} {510, L4}

\bibitem[\protect\citeauthoryear{{Ting}, {Hawkins}  \& {Rix}}{{Ting}
  et~al.}{2018}]{Ting_2018}
{Ting} Y.-S.,  {Hawkins} K.,   {Rix} H.-W.,  2018, \mn@doi [\apjl]
  {10.3847/2041-8213/aabf8e}, \href
  {https://ui.adsabs.harvard.edu/abs/2018ApJ...858L...7T} {858, L7}

\bibitem[\protect\citeauthoryear{{Vasiliev}, {Belokurov}  \&
  {Evans}}{{Vasiliev} et~al.}{2022}]{Vasiliev_2022}
{Vasiliev} E.,  {Belokurov} V.,   {Evans} N.~W.,  2022, \mn@doi [\apj]
  {10.3847/1538-4357/ac4fbc}, \href
  {https://ui.adsabs.harvard.edu/abs/2022ApJ...926..203V} {926, 203}

\bibitem[\protect\citeauthoryear{{Vincenzo}, {Spitoni}, {Calura}, {Matteucci},
  {Silva Aguirre}, {Miglio}  \& {Cescutti}}{{Vincenzo} et~al.}{2019}]{Vin2019}
{Vincenzo} F.,  {Spitoni} E.,  {Calura} F.,  {Matteucci} F.,  {Silva Aguirre}
  V.,  {Miglio} A.,   {Cescutti} G.,  2019, \mn@doi [\mnras]
  {10.1093/mnrasl/slz070}, \href
  {https://ui.adsabs.harvard.edu/abs/2019MNRAS.487L..47V} {487, L47}

\bibitem[\protect\citeauthoryear{{Xiang} \& {Rix}}{{Xiang} \&
  {Rix}}{2022}]{Xiang_2022}
{Xiang} M.,  {Rix} H.-W.,  2022, \mn@doi [\nat] {10.1038/s41586-022-04496-5},
  \href {https://ui.adsabs.harvard.edu/abs/2022Natur.603..599X} {603, 599}

\bibitem[\protect\citeauthoryear{{Zhao}, {Zhao}, {Chu}, {Jing}  \&
  {Deng}}{{Zhao} et~al.}{2012}]{Zhao_2012}
{Zhao} G.,  {Zhao} Y.,  {Chu} Y.,  {Jing} Y.,   {Deng} L.,  2012, \mn@doi
  [Research in Astronomy and Astrophysics] {10.1088/1674-4527/12/7/002}, 12,
  723

\makeatother
\end{thebibliography}







\bsp	
\label{lastpage}
\end{document}